%&amstex
%Paper: gr-qc/9508063
%From: gmichaud@campy.physics.mcgill.ca
%Date: Thu, 31 Aug 95 12:27:54 -0400

% Proceeding of Guy Michaud for the sixth Canadian Conference on General
% Relativity and Relativistic Astrophysics

\input amstex
\documentstyle{fic}
\NoBlackBoxes

\topmatter
\title Two-Dimensional Dilaton Black Holes\endtitle
\author Guy Michaud and Robert C. Myers\\
\address Department of Physics, McGill University\\
	Montr\'eal, Qu\'ebec, Canada H3A 2T8\endaddress
\endauthor

\leftheadtext{Guy Michaud and Robert C. Myers}
\rightheadtext{Fields Institute Communications Series}

\cvol{00}
\cvolyear{1995}
\cyear{1995}

%  Math Subject Classifications
\subjclass Primary 83C57; Secondary 83C75\endsubjclass
\abstract The two-dimensional CGHS model provides
an interesting toy-model for the study of black hole evaporation.
For this model, a quantum effective action, which
incorporates Hawking radiation and backreaction, can be explicitly
constructed.
In this paper, we study a generalization of this effective action.
In our extended model, it is possible
to remove certain curvature singularities arising for the original
theory. We also find that the flux of Hawking radiation
is identical to that encountered in other two-dimensional models.

\endabstract

%Replace the following by the right volume number
\cvolyear {0000 \endgraf McGill/95-40\endgraf gr-qc/9508063}

%  \thanks will become a 1st page footnote.
%  Use \endgraf to indicate a new paragraph; a blank line or \par will
%  be recognized as an error.
\thanks Talk presented by Guy Michaud. \endgraf
This research was supported by NSERC of Canada and Fonds FCAR du Qu\'ebec.
\endthanks
\endtopmatter

\document

\head 1\enspace Introduction\endhead

In 1992, Callan, Giddings, Harvey and Strominger (CGHS) presented
an interesting two-dimensional toy-model (\cite{Callan et al.~[1992]})
for the study of black hole evaporation (\cite{Hawking [1975]}).
Much greater analytic progress can be made in studying such a
two-dimensional model because of the fewer number of degrees of freedom
and reduced complexity as compared to gravity in four dimensions.
However in two dimensions, one cannot use the Einstein action
because it leads to trivial equations of motion. In the CGHS model,
an extra scalar field, the dilaton, is included to produce
a nontrivial theory of gravity.
The ``classical" CGHS model is based on the string-inspired action:
$$
S_0 = {1\over 2\pi} \int d^2x \sqrt{-g} \left\{e^{-2\phi} [R + 4
(\nabla\phi )^2 +4\lambda^2] -{1\over2}\sum^N_{i=1} (\nabla f_i)^2\right\}\ .
\tag 1.1
$$
This theory couples the two-dimensional metric, $g_{ab}$, and the dilaton,
$\phi$, to $N$ massless scalar fields, $f_i$. It also includes a
cosmological constant, $\lambda^2$.

In fact, the general solution for this theory may be constructed
analytically (\cite{Callan et al.~[1992]}).
Usually, the solutions
are written in the conformal gauge where the metric is given
in diagonal form as:
$$
g_{\pm\pm} =0 \quad\quad g_{+-} = -{1\over 2} e^{2\rho}
$$
where we use the null coordinates defined by $x^\pm = t \pm x$.
The general solution includes many black hole solutions.  The simplest
of these is the eternal black hole, in which all of the matter fields
are set to zero. In Kruskal gauge with $\rho =\phi$
(\cite{Callan et al.~[1992]}), one has:
$$
e^{-2\phi} = e^{-2\rho} = {M\over\lambda} -\lambda^2 x^+ x^-
$$
where $M$ is the mass of the black hole as it can be seen by a computation
of the Bondi mass (\cite{Callan et al.~[1992]}).
This solution has the same causal structure as the
extended Schwarzschild black hole, with future and past spacelike
singularities. With $M=0$, one is left with the linear dilaton
vacuum. In this case, the metric describes flat two-dimensional
Minkowski space while (in appropriate coordinates) the dilaton increases
linearly in the spatial direction.

Other solutions describe
the formation of a black hole from the collapse of a shell of matter.
One example, which contains an infinitely thin shell (or shock wave)
collapsing along $x^+ = x^+_0$, has:
$$
e^{-2\phi} = e^{-2\rho} = -\lambda^2 x^+ x^- -m\,(x^+ - x^+_0)\,
\Theta (x^+ -x^+_0)\ .
$$
This solution is divided in two different regions as we see from the step
function $\Theta (x^+ -x^+_0)$.  In the first region below the infall line
($x^+ <x^+_0$), the solution corresponds to the linear dilaton vacuum.
The region above the infall line ($x^+ >x^+_0$) is a portion of
the eternal black hole with an event horizon located at
$x^- =-m/\lambda^2$.  Thus it contains the future
spacelike singularity.

\subhead 1.1 \enspace One-loop effective action \endsubhead
The previous black hole solutions are classical, and so they do not
include Hawking radiation.
In order to study the latter, we must include quantum effects.
One approach is to define the quantum theory with the
following functional integral:
$$
\Cal Z = \int\Cal D g\,\Cal D \phi\,\Cal D f_i\  e^{iS_{DG}(g,\phi)
+ iS_M(g,f_i)}
\tag 1.2
$$
where $S_{DG}$ and $S_M$ denote the pure dilaton-gravity and the matter
contributions to the action $S_0$, respectively.
Actually, we cannot perform this functional integral completely, but
we can compute the matter functional integral which is a simple
Gaussian. One finds:
$$
\int\Cal D f_i\ e^{iS_M(g,f_i)} = e^{i S_{1}(g)}
$$
where $S_1(g)$ is the Polyakov action,
which may be written in a covariant non-local form
(\cite{Polyakov [1981]}):
$$
S_{1} = -{\kappa\over 8\pi} \int d^2x\sqrt{-g}\, R {1\over\nabla^2}R
\tag 1.3
$$
where $\kappa\equiv N/12$, and $1/\nabla^2$ is the Green's function
for the scalar d'Alembertian $\nabla^2$. This action can also be written
in a local form in conformal gauge.
This one-loop contribution combined with $S_0$ yields a quantum
effective action for the CGHS theory.
Solutions of this effective action incorporate
Hawking radiation and backreaction at least to leading order in
a $1/N$ expansion for large $N$. Unfortunately, it is apparently
not possible to solve the resulting equations of motion exactly,
although some qualitative results have been produced.

\head 2\enspace Generalized model\endhead

The measure used in the
functional integral (1.2) is not uniquely defined,
and one can define alternative
theories with different measures. Explicitly, this means that
we can build new theories by adding local, covariant counterterms to
the Polyakov action $S_{1}$.  Here, we exploit
this freedom to build a generalization of the previous quantum effective
action, and look at the effects of the new interactions
on black hole physics. Our extended action will be
$S =S_0 + S_1 + S_2 + S_3$, where $S_0$ is the
classical CGHS action (1.1) and $S_1$ is the
Polyakov action (1.3).  The remaining terms are:
$$
\aligned
S_2 &= -{\kappa\over 8\pi} \int d^2x \sqrt{-g}
\left[\alpha\phi R +\beta (\nabla\phi )^2\right]\\
S_3 &= -{\kappa\over 8\pi} \int d^2x \sqrt{-g}
\sum^{K}_{n=2}\left[a_n\phi^n R + b_n\phi^{n-1}(\nabla\phi )^2\right]
\endaligned
$$
where $\alpha$, $\beta$, $a_n$ and $b_n$ are coupling constants,
and $K\ge2$ is some integer. One of our objectives
in selecting the coupling constants will be to produce an exactly
soluble theory. A condition which will achieve this result is
requiring the simple current equation (\cite{Russo et al.~[1992],
Kazama et al.~[1995]}):
$$
\partial_+\partial_- (\rho -\phi )=0\ .
$$
The difference between $T_{+-}$ and the dilaton equation of motion
yields the above if we set:
$$
\aligned
\beta\, &= 4-2\alpha \\
b_n &= -{2n}\, a_n\ .
\endaligned
$$
Note that further setting $\alpha=2$ and $a_n=0$
(and hence $\beta=0=b_n$) produces the
model of Russo, Susskind and Thorlacius (Russo et al.[1992]).

\subhead 2.1\enspace Liouville theory \endsubhead
With the above simple current conditions and in conformal gauge,
our action $S$ is simplified by performing the field redefinition:
$$
\aligned
\chi &= \rho -{\alpha\over 4}\phi + {1\over\kappa} e^{-2\phi}
-{1\over 4}\sum^K_{n=2} a_n\phi^n \\
\Omega &= \left( 1-{\alpha\over 4}\right)\phi + {1\over\kappa} e^{-2\phi}
-{1\over 4}\sum^K_{n=2} a_n\phi^n\ .
\endaligned
$$
With these new fields, the action takes a Liouville form:
$$
S ={1\over\pi} \int d^2x \left\{-\kappa\;\partial_+\chi\partial_-\chi
    +\kappa\;\partial_+\Omega\partial_-\Omega
    +\lambda^2 e^{2 (\chi -\Omega )} + {1\over 2}
    \sum^N_{i=1} \partial_+f_i\partial_-f_i\right\}\ .
$$
In terms of the fields $\chi$ and $\Omega$, we can write the Ricci
curvature scalar as:
$$
\aligned
R &= 8e^{-2\rho} \partial_+\partial_-\rho \\
  &= 8e^{-2\rho (\chi ,\Omega)} {1\over\Omega^\prime}\left[
    \partial_+\partial_-\chi -{\Omega^{\prime\prime}\over\Omega^\prime}
\partial_+\Omega\partial_-\Omega\right]
\endaligned
$$
where $\rho (\chi ,\Omega )$ is the conformal factor as an implicit function
of the new fields and
the prime ($^\prime$) denotes a derivative with respect to the field $\phi$.
Hence the curvature diverges whenever $\Omega^\prime$ vanishes.
Such extrema of the function $\Omega (\phi )$ lead to timelike
singularities, even for the vacuum solution in which the dilaton
always increases along the spatial direction.  However,
in the extended model,
we can avoid these singularities by properly constraining the coupling
constants.  For example, $\Omega (\phi )$ has no extrema
if $\alpha>4$, $a_n>0$ for odd $n$, and $a_n=0$ for even $n$.

\subhead 2.2\enspace Solution\endsubhead
We can further simplify the solutions with another field rede\-fi\-ni\-tion:
$$
\aligned
U &= {\kappa\over 2} (\chi +\Omega )\\
V &= \chi - \Omega\ .
\endaligned
$$
Then the metric equations of motion become:
$$
\aligned
T_{\pm\pm} &= -2\partial_\pm U\partial_\pm V
 + {\kappa\over 2}\partial^2_\pm V + \partial_\pm ^2 U
 + {1\over 2}\sum^N_{i=1}(\partial_\pm f_i)^2+t_\pm (\sigma^\pm ) =0 \\
T_{+-} &= -{\kappa\over 2}\partial_+\partial_- V -\partial_+\partial_-U
-\lambda^2 e^{2V} =0
\endaligned
$$
where the functions $t_\pm$ arise because of zero-mode
ambiguities in defining the matter Green's function in the Polyakov
action (1.3). We consider the
solution for a collapsing shock wave for which
the matter configuration is:
$$
T^f_{++} = {1\over 2}\sum^N_{i=1}(\partial_+ f_i)^2
= m\,\delta (\sigma^+ -\sigma^+_0) \qquad
T^f_{--}= {1\over 2}\sum^N_{i=1}(\partial_- f_i)^2 =0
$$
where $m$ is the amplitude of the shock wave. We also choose
$t_\pm$-functions to have the form (\cite{Russo et al.~[1992]}):
$$
t_\pm (x^\pm ) = -{\kappa\over 4} {1\over (x^\pm )^2}\ .
$$
Now before presenting the solution, we transform to
asympto\-ti\-cally Minkowskian coordinates for which
at large radius $ds^2\simeq-d\sigma^+d\sigma^-$.
With these choices, the solution is:
$$
\aligned
U &= e^{\lambda (\sigma^+ -\sigma^-)} -{m\over\lambda}
  \left( e^{\lambda (\sigma^+ -\sigma^+_0)} -1\right)
  \Theta (\sigma^+ -\sigma^+_0)
  -{\kappa\over 4}\ln \left[1 + {m\over\lambda} e^{\lambda\sigma^-}\right]\\
V &= {\lambda\over 2} (\sigma^+ -\sigma^-)\ .
\endaligned
\tag 2.1
$$
This solution is the analogue of the classical shock wave.  However, the
above solution now includes the effects of Hawking radiation.
The black hole evaporation can be examined by looking at the
evolution of the Bondi mass.

\head 3\enspace Bondi mass\endhead

The Bondi mass is defined on the future null infinity ${\Cal J}^+_R$ and
in the asymptotically Minkowskian coordinates, $\sigma^\pm$.  It is
expressed in terms of the linear variations $\delta T_{ab}$ of the
stress-energy tensor around some reference solution.  As a reference,
we take the ``vacuum solution'' which is given by setting
$m=0$ in the shock wave solution (2.1).
Thus, the Bondi mass (\cite{de Alwis [1992]}):
$$
M (\sigma^-) = -\int^{{\Cal J}^+_R} d\sigma^+ (\delta T_{++} + \delta T_{+-})
$$
for the shock wave solution is:
$$
M (\sigma^-) = m - {\lambda\over 4}\kappa \left\{
\ln \left[ 1+ {m\over\lambda} e^{\lambda\sigma^-} \right]
   + {m\over \lambda e^{-\lambda\sigma^-} + m}\right\}\ .
$$
The flux of Hawking radiation may be
obtained by differentiating this expression with respect to $\sigma^-$.
Note that our results are completely independent of any of the
coupling constants $\alpha$ and $a_n$. Hence the radiation
in these solutions is essentially the same as
in other two-dimensional models (\cite{Callan et al.~[1992],
Russo et al.~[1992]}).
The radiation goes to zero in the far past
$\sigma^-\rightarrow -\infty$, while
it approaches a constant in the far future $\sigma^-\rightarrow\infty$.
The latter constant flux means that the black holes never stop Hawking
radiating even when their mass reaches zero! Thus our generalized
model does not avoid this unphysical behavior found in other models.

\head Conclusion \endhead

We have presented a new generalization for the effective
quantum action derived by CGHS for
the study of Hawking radiation in two dimensions. We have
shown that our generalization gives the possibility of removing the
timelike curvature singularity arising in other models. Hence we may
avoid the related boundary
problems (\cite{Das and Mukherji [1994], Russo et al.~[1993],
Strominger and Thorlacius [1994]}).  Also, this model
is exactly solvable in the sense that we can
find the general solution analytically. By studying
the solution for the collapsing shock wave, we found that black hole
evaporation proceeds in essentially the same way as
in other dilaton gravity models. In
particular, the Hawking radiation in our generalized model
never stops! Thus we must conclude that the physics of this
generalized theory remains
incomplete, and we are still unable to determine the end-state of
the black hole evaporation.

\refstyle{B}


\Refs\nofrills{References}

\ref
\by	Callan, C.G., Giddings, S.B., Harvey, J.A., and Strominger, A.
\paper	Evanescent black holes,
\jour	Phys. Rev.
\vol	D46 \yr 1992 \pages R1005--R1009
\endref

\ref
\by	Das, S.R., and Mukherji, S.
\paper  Boundary dynamics in dilaton gravity,
\jour	Mod. Phys. Lett.
\vol	A9 \yr 1994 \pages 3105--3118
\endref

\ref
\by	de Alwis, S.P.
\paper	Quantum black holes in two dimensions,
\jour	Phys. Rev.
\vol	D46 \yr 1992 \pages 5429--5438
\endref

\ref
\by 	Hawking, S.W.
\paper	Particle creation by black holes,
\jour	Comm. Math. Phys.
\vol	43 \yr 1975 \pages 199--220
\endref

\ref
\by	Kazama, Y., Satoh, Y., and Tsuchiya, A.
\paper	A Unified approach to solvable models of dilaton gravity
in two-dimensions based on symmetry,
\jour	Phys. Rev.
\vol	D51 \yr 1995 \pages 4265--4276
\endref

\ref
\by	Polyakov, A.M.
\paper	Quantum geometry of bosonic strings,
\jour	Phys. Lett.
\vol	B103 \yr 1981 \pages207--210
\endref

\ref
\by	Russo, J.G., Susskind, L., and Thorlacius, L.
\paper	End point of Hawking radiation,
\jour	Phys. Rev.
\vol	D46 \yr 1992 \pages 3444--3449
\endref

\ref
\by     ------
\paper  Cosmic censorship in two-dimensional gravity,
\jour   Phys. Rev.
\vol    D47 \yr 1993 \pages 533--539
\endref

\ref
\by     Strominger, A., and Thorlacius, L.
\paper  Conformally invariant boundary conditions for dilaton gravity,
\jour   Phys. Rev.
\vol    D50 \yr 1994 \pages 5177--5187
\endref

\endRefs
\enddocument